%% file: main.tex
\documentclass[sigconf,authorversion,nonacm]{acmart}
\input{document/packages}
\setcopyright{acmlicensed}
\copyrightyear{2018}
\acmYear{2018}
\acmDOI{XXXXXXX.XXXXXXX}
\acmConference[Conference acronym 'XX]{Make sure to enter the correct
  conference title from your rights confirmation email}{June 03--05,
  2018}{Woodstock, NY}
\acmISBN{978-1-4503-XXXX-X/2018/06}

\begin{document}
\input{document/header}

\begin{abstract}
\input{document/abstract}
\end{abstract}

\input{document/meta}
\input{document/teaser}

\received{\today}

\maketitle
\input{sections/index}

%\begin{acks}
%ACKNOWLEDGEMENTS GO HERE LATER
%\end{acks}

\newpage
\bibliographystyle{ACM-Reference-Format}
\bibliography{references}

\input{appendix/index}

\end{document}

%% file: document/packages.tex
\usepackage{geometry}
\usepackage{xcolor}
\usepackage{subcaption}
\usepackage{mwe}
\usepackage{listings}

\definecolor{codegreen}{rgb}{0,0.6,0}
\definecolor{codegray}{rgb}{0.5,0.5,0.5}
\definecolor{codepurple}{rgb}{0.58,0,0.82}
\definecolor{backcolour}{rgb}{0.97,0.97,0.97}
\definecolor{keywordblue}{rgb}{0.13, 0.13, 1}
\definecolor{attributered}{rgb}{0.64,0.08,0.08}

\lstdefinestyle{mystyle}{
    backgroundcolor=\color{backcolour},
    commentstyle=\color{codegreen},
    numberstyle=\tiny\color{codegray},
    stringstyle=\color{codepurple},
    basicstyle=\ttfamily\small,
    breakatwhitespace=false,
    breaklines=true,
    captionpos=b,
    keepspaces=true,
    showspaces=false,
    showstringspaces=false,
    showtabs=false,
    tabsize=2,
    frame=single,
    framerule=0pt,
    rulecolor=\color{black}
}

\lstdefinelanguage{VOIXHTML}{
    language=html,
    morekeywords={context, tool, prop, name, description, required, type, @call, v-for, return}, % VOIX elements
    keywordstyle=\color{keywordblue}\bfseries,
    sensitive=false,
    morecomment=[s]{},
    morestring=[b]",
    morestring=[b]',
}

\lstdefinelanguage{JavaScript}{
  keywords={break, case, catch, continue, const, let, var, reactive, function, return, if, else, for, switch, try, new, true, false, null, typeof, await, async},
  keywordstyle=\color{keywordblue}\bfseries,
  ndkeywords={document, window, event, detail, push},
  ndkeywordstyle=\color{teal},
  identifierstyle=\color{black},
  sensitive=true,
  comment=[l]{//},
  morecomment=[s]{/*}{*/},
  commentstyle=\color{codegreen}\itshape,
  stringstyle=\color{codepurple},
  morestring=[b]',
  morestring=[b]"
}

\usepackage{hyperref}

\makeatletter
\newcommand\Autoref[1]{\@first@ref#1,@}
\def\@throw@dot#1.#2@{#1}% discard everything after the dot
\def\@set@refname#1{%    % set \@refname to autoefname+s using \getrefbykeydefault
    \edef\@tmp{\getrefbykeydefault{#1}{anchor}{}}%
    \xdef\@tmp{\expandafter\@throw@dot\@tmp.@}%
    \ltx@IfUndefined{\@tmp autorefnameplural}%
         {\def\@refname{\@nameuse{\@tmp autorefname}s}}%
         {\def\@refname{\@nameuse{\@tmp autorefnameplural}}}%
}
\def\@first@ref#1,#2{%
  \ifx#2@\autoref{#1}\let\@nextref\@gobble% only one ref, revert to normal \autoref
  \else%
    \@set@refname{#1}%  set \@refname to autoref name
    \@refname~\ref{#1}% add autoefname and first reference
    \let\@nextref\@next@ref% push processing to \@next@ref
  \fi%
  \@nextref#2%
}
\def\@next@ref#1,#2{%
   \ifx#2@ and~\ref{#1}\let\@nextref\@gobble% at end: print and+\ref and stop
   \else, \ref{#1}% print  ,+\ref and continue
   \fi%
   \@nextref#2%
}

\usepackage{csquotes}
\MakeOuterQuote{"}

\usepackage{enumitem}
\newlist{requirements}{enumerate}{1}

\setlist[requirements,1]{label=\textbf{R\arabic*}, ref=R\arabic*}

\usepackage{dblfloatfix}
\usepackage{multirow}
\usepackage{tablefootnote}

\usepackage{array}

%% file: document/header.tex
\title[Building the Web for Agents: A Declarative Framework for Agent–Web Interaction.]{Building the Web for Agents:\\A Declarative Framework for Agent–Web Interaction.}

% AUTHORS
\author{Sven Schultze}
\email{sven.schultze@tu-darmstadt.de}
\authornote{Authors contributed equally to this research.}
\affiliation{
  \institution{Technical University of Darmstadt}
  \city{Darmstadt}
  \state{Hessen}
  \country{Germany}
}

\author{Meike Kietzmann}
\email{meike.kietzmann@tu-darmstadt.de}
\authornotemark[1]
\affiliation{%
  \institution{Technical University of Darmstadt}
  \city{Darmstadt}
  \state{Hessen}
  \country{Germany}
}

\author{Nils Lucas Schoenfeld}
\email{nilslucas.schoenfeld@tu-darmstadt.de}
\authornotemark[1]
\affiliation{
  \institution{Technical University of Darmstadt}
  \city{Darmstadt}
  \state{Hessen}
  \country{Germany}
}

\author{Ruth Stock-Homburg}
\email{rsh@bwl.tu-darmstadt.de}
\affiliation{
  \institution{Technical University of Darmstadt}
  \city{Darmstadt}
  \state{Hessen}
  \country{Germany}
}

\renewcommand{\shortauthors}{Schultze et al.}

%% file: document/abstract.tex
The increasing deployment of autonomous AI agents on the web is hampered by a fundamental misalignment: agents must infer affordances from human-oriented user interfaces, leading to brittle, inefficient, and insecure interactions. To address this, we introduce VOIX, a web-native framework that enables websites to expose reliable, auditable, and privacy-preserving capabilities for AI agents through simple, declarative HTML elements. VOIX introduces <tool> and <context> tags, allowing developers to explicitly define available actions and relevant state, thereby creating a clear, machine-readable contract for agent behavior. This approach shifts control to the website developer while preserving user privacy by disconnecting the conversational interactions from the website. We evaluated the framework's practicality, learnability, and expressiveness in a three-day hackathon study with 16 developers. The results demonstrate that participants, regardless of prior experience, were able to rapidly build diverse and functional agent-enabled web applications. Ultimately, this work provides a foundational mechanism for realizing the Agentic Web, enabling a future of seamless and secure human-AI collaboration on the web.

%% file: document/meta.tex
\begin{CCSXML}
<ccs2012>
   <concept>
       <concept_id>10003120.10003121.10003124.10010868</concept_id>
       <concept_desc>Human-centered computing~Web-based interaction</concept_desc>
       <concept_significance>500</concept_significance>
       </concept>
   <concept>
       <concept_id>10010147.10010178.10010219.10010221</concept_id>
       <concept_desc>Computing methodologies~Intelligent agents</concept_desc>
       <concept_significance>500</concept_significance>
       </concept>
   <concept>
       <concept_id>10002951.10003260.10003300</concept_id>
       <concept_desc>Information systems~Web interfaces</concept_desc>
       <concept_significance>300</concept_significance>
       </concept>
   <concept>
       <concept_id>10002978.10003029.10011150</concept_id>
       <concept_desc>Security and privacy~Privacy protections</concept_desc>
       <concept_significance>300</concept_significance>
       </concept>
   <concept>
       <concept_id>10003120.10003121.10003122.10003334</concept_id>
       <concept_desc>Human-centered computing~User studies</concept_desc>
       <concept_significance>100</concept_significance>
       </concept>
 </ccs2012>
\end{CCSXML}

\ccsdesc[500]{Human-centered computing~Web-based interaction}
\ccsdesc[500]{Computing methodologies~Intelligent agents}
\ccsdesc[300]{Information systems~Web interfaces}
\ccsdesc[300]{Security and privacy~Privacy protections}
\ccsdesc[100]{Human-centered computing~User studies}
\keywords{Agentic Web, AI Agents, Web Agents, Large Language Models, Machine-Readable Web, Multimodal Interaction, Developer Experience, Web Standards, Privacy, Decentralization}

%% file: document/teaser.tex
%   \includegraphics[width=\textwidth]{figures/sampleteaser.pdf}

%% file: sections/index.tex
\input{sections/introduction}
\input{sections/related-work}
\input{sections/voix}
\input{sections/hackathon}

\input{sections/discussion}
\input{sections/conclusion}

%% file: sections/introduction.tex
\section{Introduction}
The past years have seen rapid progress in large language models (LLMs) and their integration into interactive systems. Increasingly, these models are being deployed as autonomous or semi-autonomous agents capable of acting on behalf of users in complex environments such as the web. Research like WebArena \cite{zhou_webarena_2024} has benchmarked such agents in realistic, long-horizon tasks, while commercial systems (e.g., Claude for Chrome, Perplexity Comet, Gemini in Chrome) demonstrate growing interest in agent-mediated browsing. Yet, the integration of conversational agents into the existing web ecosystem remains fundamentally misaligned with the current architecture of the web. 

Today’s web is designed primarily for human consumption. Agents must infer available actions by scraping HTML, heuristically parsing Document Object Models (DOMs) or even analyzing rendered screenshots. With these ad hoc practices, even minor state changes can disrupt agents’ workflow. Agents are inefficient, as they have to repeatedly rediscover affordances. In addition, they are insecure, since unintended operations or unauthorized data access cannot be ruled out: Sensitive, personal, or proprietary information embedded in the web page, such as private messages, financial data, or user details, could be shared without the user's explicit consent.

Empirical evaluations show that browsing-only agents underperform human users on realistic tasks, and that augmenting browsing with machine-native APIs or hybrid browsing+API access yields substantial gains in completion rates and efficiency \cite{zhou_webarena_2024, he_webvoyager_2024}. This highlights a deeper structural challenge: While the web has evolved rich standards for human interaction, it lacks equivalent, machine-native affordances for agents. Without explicit, declarative mechanisms to declare actions and state, AI developers are forced to retrofit agent capabilities onto interfaces built for humans, leading to fragile integrations, long trajectories, and limited scalability. 

Finally, the current paradigm strips website developers of control over the user experience on their own pages. When an external agent scrapes a site, it bypasses the carefully crafted workflows and interaction patterns designed by the developer. The agent provider, not the site owner, unilaterally decides how to interpret and interact with the page's functionality. This lack of an explicit, machine-readable contract leaves developers unable to communicate their site's capabilities, define safe actions, or protect sensitive data, creating an unpredictable and unstable environment for both agents and the websites they navigate. Recent position papers \cite{yang_agentic_2025, lu_build_2025} converge on the need for an Agentic Web. In this web, machine-readable, standardized affordances are first-class citizens and agents can operate without reverse engineering user interfaces built for humans.

To this end, we introduce VOIX, a web-native framework that embodies this principle through a simple, declarative substrate for robust, privacy-preserving agent–web interaction. VOIX enables websites to explicitly declare actions and relevant state in a way that is equally accessible to autonomous agents, reducing the need for brittle inference from human-oriented user interfaces (UIs) and lowering the development barrier for rich, multimodal experiences. This shifts agent–web interaction from a model where external providers unilaterally interpret a site’s DOM, to one where the site developer defines an explicit, auditable contract for agent behavior.

To validate the practicality and accessibility of VOIX, we conducted a three-day hackathon study with 16 developers. The results demonstrate that VOIX can be rapidly adopted to build diverse agent-enabled web applications, providing initial empirical evidence of its learnability and expressiveness.

Our work advances the research discourse on agent-web interaction by operationalizing the Agentic Web paradigm, which has been described in the literature as a vision but has not yet been realized in practice \cite{lu_build_2025, yang_agentic_2025}. While prior work has identified machine-readable affordances as a necessary foundation for agentic web interaction, VOIX provides the first concrete implementation, translating abstract calls for Agentic Web Interfaces into an empirically validated mechanism.

Next, we offer a normative model for trust and governance in agentic systems. In a landscape where current architectures either centralize control with inference providers or with site operators, VOIX formalizes an alternative model that distributes responsibility through explicit, auditable contracts of interaction among web developers, inference providers, and end-users. In doing so, VOIX contributes to broader theoretical debates on safety, privacy, and accountability in human-AI collaboration.

%% file: sections/related-work.tex
\section{Related Work}
The design of VOIX was informed by research on web agents, multimodal interaction, and emerging calls to standardize agentic interfaces for the web. We review these strands of prior work to summarize the state of the art, surface core limitations, and synthesize requirements for VOIX.

\subsection{Web Agents and Environments}
Recent position and survey papers argue that the web is entering an “Agentic Web” era, in which autonomous agents act on users’ behalf through goal-directed interaction, demanding new protocols and semantics beyond human-only UIs \cite{yang_agentic_2025}. Complementing this vision, \citet{lu_build_2025} contend that forcing agents to adapt to human-facing DOM and screenshots is misaligned with agent capabilities and safety. They advocate Agentic Web Interfaces: standardized, machine-native affordances that are safe, transparent, and efficient for agents to consume.

Realistic, reproducible environments have clarified the limits of current agents and the opportunity space for machine-native interfaces. WebArena assembles fully functional websites spanning common domains and shows that even strong models underperform humans on long-horizon tasks \cite{zhou_webarena_2024}. WebVoyager demonstrates an LMM-driven, end-to-end agent for real sites and introduces an evaluation protocol, substantially outperforming text-only baselines \cite{he_webvoyager_2024}.

Crucially, exposing machine-native endpoints materially improves agent success. \citet{song_beyond_2025} compare browsing-only, API-only, and hybrid agents on WebArena, finding Hybrid (browsing + APIs) achieves improvements over browsing alone, and that API quality strongly moderates outcomes. \citet{reddy_infogent_2024} further demonstrate, for information aggregation across multiple sites, a modular agent with Navigator–Extractor–Aggregator components, showing gains under both direct API-driven access and interactive visual access.

A recent survey synthesizes components, data, and evaluation practices for LLM-based graphical user interface (GUI) agents, mapping systems across perception, grounding, planning, and action spaces \cite{zhang_large_2025}. In mobile, \citet{zhang_appagent_2025} present AppAgent, which interacts with smartphone apps via a simplified action space (tap, long-press, swipe, text, back, exit) and a two-phase learning process (exploration or observing demonstrations to build reference documentation). Empirically, AppAgent’s simplified, machine-readable affordances and documentation substantially improve success over raw coordinates and no-doc baselines.

\subsection{Multimodal Interaction}
Decades of human–computer interaction research show that interaction is more powerful when combining modalities such as natural language and direct manipulation. Oviatt’s foundational work on multimodal systems \cite{oviatt_ten_1999} demonstrates that pairing speech or text with pointing or selection reduces ambiguity, increases robustness, and is strongly preferred by users for spatial or object-specific tasks. Recent LLM-powered systems confirm this from two complementary directions: DirectGPT \cite{masson_directgpt_2024} enriches chat-based interaction with graphical cues such as selection and pointing, while ReactGenie \cite{yang_reactgenie_2024} enriches GUIs with voice input interpreted by an LLM. Building on this, GenieWizard simulates user personas and dialogues to elicit multimodal commands, uses a zero-shot parser plus abstract interpretation to identify missing APIs, and helps developers close coverage gaps \cite{yang_geniewizard_2025}. Earlier, Geno showed that developer-side augmentation can retrofit multimodality onto existing web apps \cite{sarmah_geno_2020}. \citet{wen_exploring_2025} show that adding visual prompts and direct manipulation to text improves disambiguation in visualization without sacrificing efficiency. On mobile text correction, Tap\&Say integrates touch location into an LLM’s context, outperforming state-of-the-art text correction baselines and commercial voice access while reducing keystrokes and effort \cite{zhao_tapsay_2025}.

Together, these results show that restricting agent interaction to a chat-only paradigm omits proven opportunities for richer, more efficient interaction. Yet, multimodal systems have historically been more difficult to implement than single-modality interfaces. To avoid user confusion, the feature sets across modalities must remain aligned, and maintaining this parity significantly increases development complexity. Non-visual cues are inherently harder for users to discover because they lack persistent visibility. On the web, these challenges intersect with the existing limitations faced by agents: without a standard interface for exposing capabilities and state, developers must retrofit multimodal support onto human-oriented UIs, compounding both the engineering burden and the brittleness of agent behavior.

\subsection{Human-in-the-Loop Agentic Systems}
The growth of agent capabilities raises concerns about misalignment, deception, unsafe exploration, security, and broader societal impacts \cite{chan_harms_2023}. Systems increasingly add scaffolds for oversight and control. Magentic-UI proposes an extensible, multi-agent interface with co-planning and co-tasking, action approval, answer verification, memory, and multi-tasking \cite{mozannar_magentic-ui_2025}. Results across agentic benchmarks and user studies suggest that lightweight human participation improves safety and outcomes.

A practical friction is that approval processes are often highly individual to the use cases and application domains: An online banking application will likely require different approval processes than a static news article. This emphasizes a central argument that the transition into the Agentic Web should involve website maintainers as central stakeholders of its architecture, instead of agent providers attempting to infer these complex, domain-specific safety requirements from the outside. 

\subsection{Positioning VOIX}
This prior work reveals the need for a standardized, machine-native protocol for the web that leverages powerful multimodal patterns, yet it also highlights the critical challenges of agent brittleness, implementation complexity, and the need for developer-defined safety and control. To address these challenges, we propose VOIX, a web-native framework for agentic interaction. The design of VOIX is guided by a set of core principles derived from architectural principles for building the Agentic Web \cite{lu_build_2025}, and foundational research on effective multimodal interaction \cite{oviatt_ten_1999, oviatt_paradigm_2015}.

\begin{requirements}
    \item \label{req:privacy} \textbf{Privacy and Safety}: Control over privacy and safety must be placed in the hands of developers and users. Developers must be able to define what data is exposed and which tools are safety-critical, while users must retain control over their conversational data and choice of LLM provider. This counters the significant privacy risks of sending full page content to third-party services.
    \item \label{req:representation} \textbf{Optimal Representation}: The framework must provide an efficient, machine-readable representation of a website's affordances. This representation must be dynamically scoped, containing only the necessary information for the agent to optimally solve its tasks, excluding all other data.
    \item \label{req:hosting} \textbf{Efficient to Host}: The architecture must not place the computational and financial burden of LLM inference on the website owner. By operating on the client-side, the framework respects the decentralized nature of the web and removes a major barrier of widespread adoption.
    \item \label{req:standardized} \textbf{Standardized and Developer-Friendly}: The framework must be built on a universal standard using familiar web patterns. This addresses the need for easy developer adoption and aligns with the call for a standardized Agentic Web Interfaces that is compatible across various agents and websites.
    \item \label{req:expressive} \textbf{Expressive}: The framework must also be expressive enough to model the key patterns of effective multimodal interaction as identified by Oviatt \cite{oviatt_ten_1999, oviatt_paradigm_2015}. This is a critical requirement for moving beyond simple feature substitution towards truly synergistic and flexible interfaces. Specifically, the framework should handle both simultaneous and sequential integration of inputs, abstract and high-level user inputs, and the ability to implement both complementary and redundant actions.
\end{requirements}

%% file: sections/voix.tex
\section{VOIX: A Web-Native Interface for Agents}\label{sec:voix}

\begin{figure*}[bt]
  \centering
  % Left column (two stacked)
  \begin{minipage}[c]{0.51\linewidth}
    \begin{subfigure}[t]{\linewidth}
      \includegraphics[width=\linewidth]{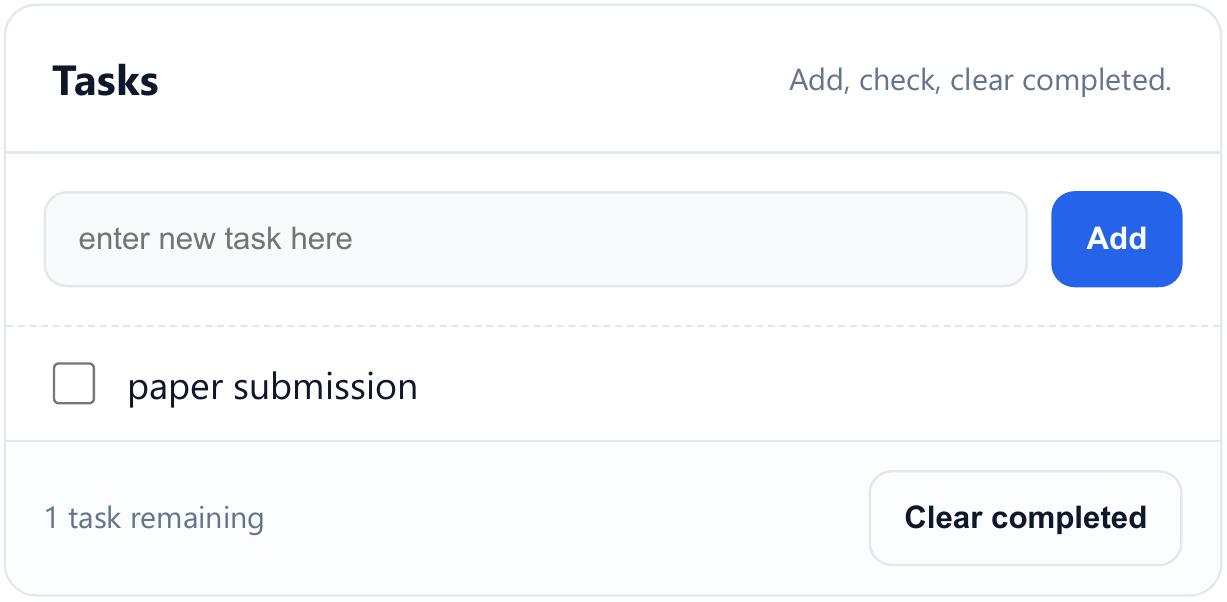}
      \subcaption{Example Web App: Todo list}
      \label{fig:example-gui}
    \end{subfigure}
    \centering

    \vspace{20pt}

    \begin{subfigure}[t]{\linewidth}
      \fbox{\includegraphics[width=0.97\linewidth]{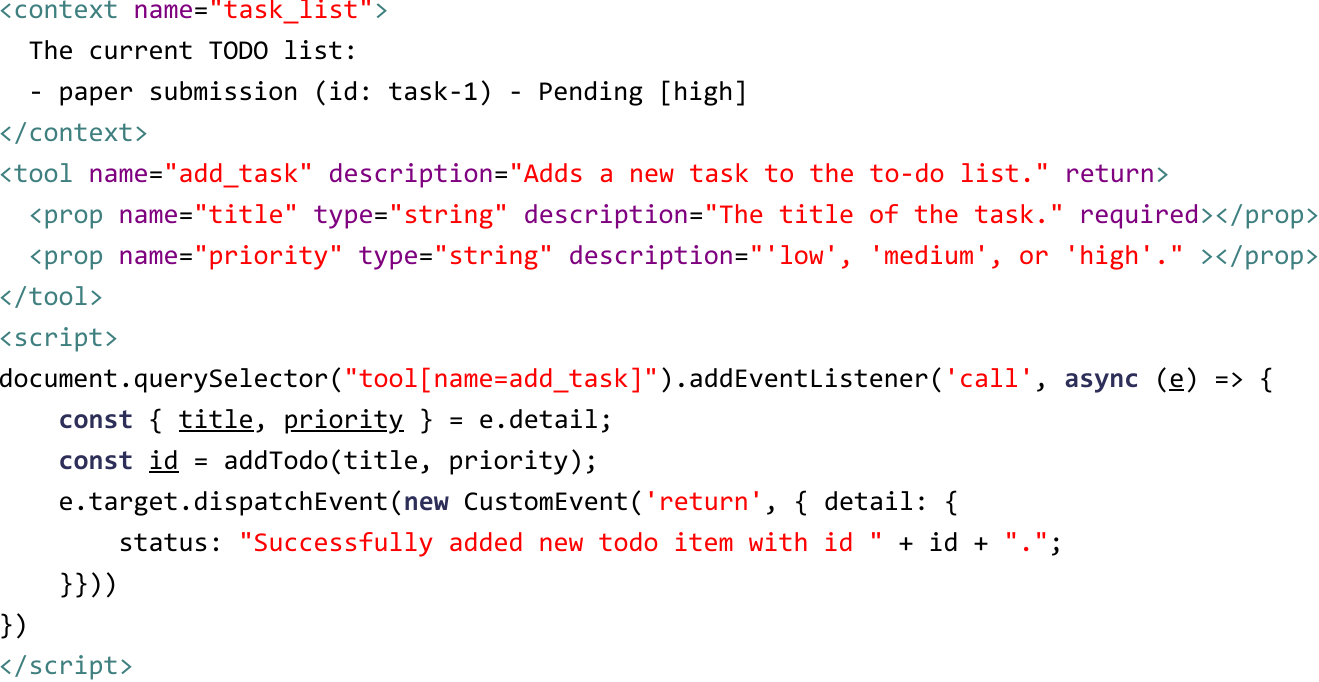}}
      \subcaption{Embedded VOIX Elements in the Web App}
      \label{fig:embedded-voix-elements}
    \end{subfigure}
  \end{minipage}\hfill
  % Right column (one tall)
  \begin{minipage}[c]{0.445\linewidth}
    \centering
    \begin{subfigure}[b]{\linewidth}
      \includegraphics[width=\linewidth]{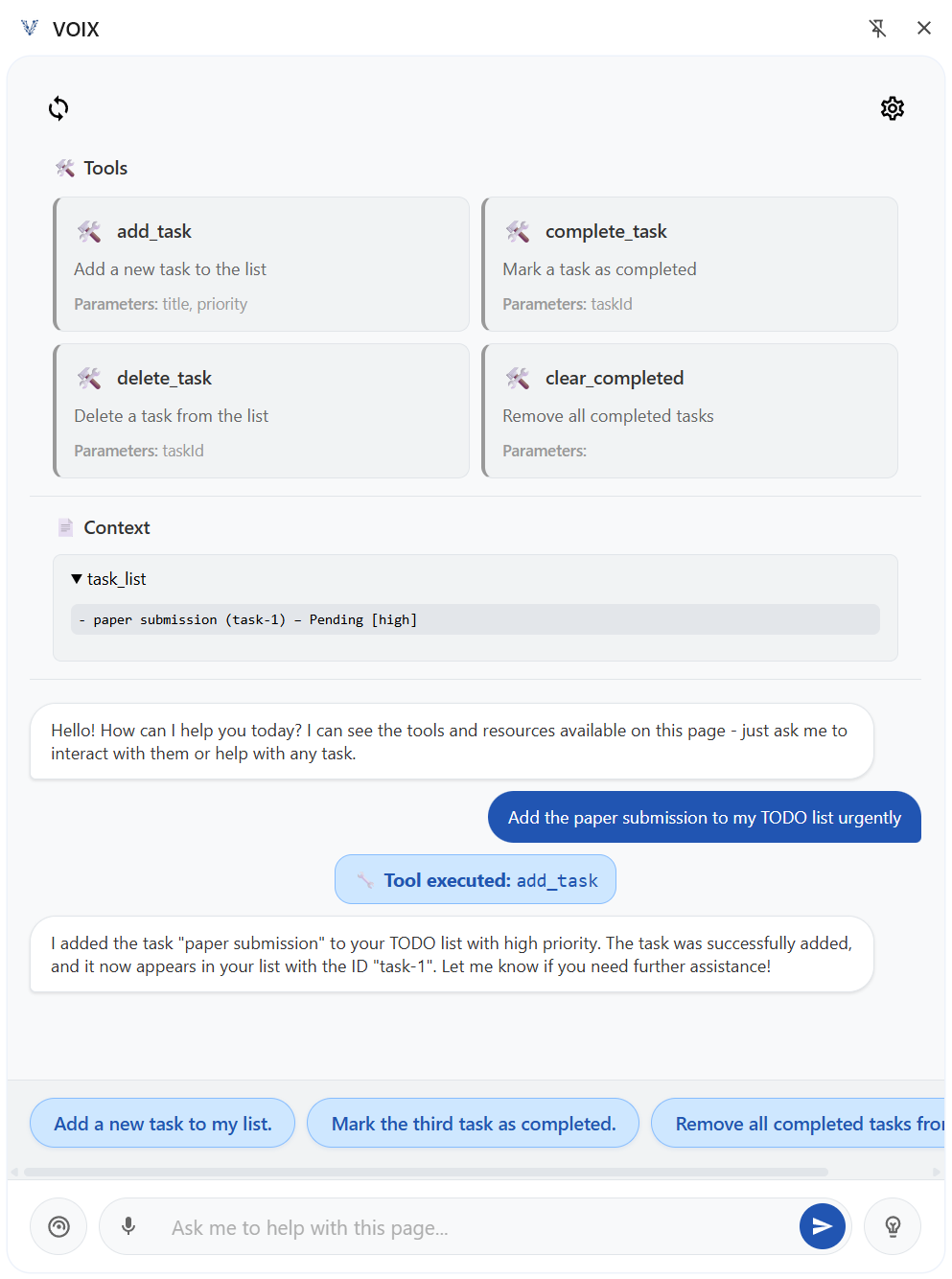}
      \subcaption{VOIX Reference Chrome Extension Sidepanel}
      \label{fig:voix-sidepanel}
    \end{subfigure}
  \end{minipage}

  \caption{Integration of VOIX into a task management web application. (a) The human-facing interface allows users to add, complete, and clear tasks. (b) The application embeds VOIX-HTML elements  in its markup, declaratively exposing state and invokable actions to match or complement these features. (c) The VOIX reference Chrome extension automatically discovers the declarations and surfaces them in a side panel, where an LLM-powered agent can reason over available tools and execute actions based on voice or chat user input.}
  \label{fig:three-subfig}
\end{figure*}

VOIX contributes a concrete, web-native mechanism that makes site capabilities and state discoverable and invokable by agents through declarative, typed semantics. VOIX introduces two new HTML elements: \texttt{<tool>} exposes actions with names, parameter types, and natural language descriptions; \texttt{<context>} exposes task-relevant state. VOIX aims to: (1) improve agent reliability and efficiency by eliminating affordance inference; (2) preserve the web's decentralization and backward compatibility; (3) provide human control, privacy, and transparency by design; and (4) remain model- and provider-agnostic so multiple agent stacks can interoperate.

To this end, VOIX defines a clear architectural model that decouples the website's functional capabilities from the agent's reasoning and execution, creating a standardized interface that prioritizes security, privacy, and decentralization. The VOIX architecture distributes responsibilities among three distinct stakeholders:

\paragraph{The Website} The web page acts as the authoritative source of its own capabilities. Its responsibility is to declare a set of invokable tools and expose relevant application context using the HTML markup. It is also responsible for implementing the business logic that executes a tool call and, if necessary, returns a result. Thereby, the websites declares the contract in which an LLM is allowed to interact with it. This happens alongside the development of the user interface, allowing developers integrate their existing application logic seamlessly and to benefit from the entire ecosystem of modern frameworks like React and Vue or server-side like Laravel without the need to install new packages.

\paragraph{The Browser Agent} The browser serves as the intermediary, decoupling the website from the Inference Provider. Its primary functions are to: (a) scan the website to discover and catalog all declared \texttt{<tool>} and \texttt{<context>} elements; (b) present this catalog to the agent for reasoning; and (c) dispatch events to trigger tool execution on the web page when instructed by the agent. This architectural role is designed to be implementation-agnostic and can be realized through various means, including open-source extensions that support multiple inference providers, enterprise-specific modules with enhanced security and integration, or native browser integrations. Our reference implementation, for instance, is a Chrome extension that demonstrates this open, provider-agnostic model.

\paragraph{The Inference Provider} The LLM, which can be cloud-based or a local model hosted by the user, is the decision-making component. It receives the catalog of tools and context from the browser agent and, based on the user's natural language objective, selects the appropriate tool and parameters for execution. Its interaction is with the structured declarations, not the visual UI.\\

This decoupled architecture enables a diverse ecosystem of interaction models, each with distinct trust and data-flow characteristics. The framework's flexibility supports a wide range of scenarios. For instance, a user could opt for a fully sovereign, private execution by combining an open-source browser extension with a locally hosted LLM, ensuring no data leaves their personal machine. Alternatively, a user could connect to a powerful cloud-based model like OpenAI's GPT or Anthropic's Claude for maximum capability. This flexibility also extends to more controlled environments. In a corporate setting, an enterprise could deploy a proprietary extension with Single Sign-On to securely interface with an internal LLM. It also accommodates native integration, where a browser might embed support directly into its product and connect to its own inference provider to offer a seamless, vertically integrated user experience.

\subsection{Trust Boundaries}
The architecture of VOIX creates clear trust boundaries by keeping the user's conversation private while giving the website and the user strict control over data sharing. When a user gives a command, their words are sent directly from the Browser Agent to their chosen Inference Provider, ensuring the website never sees the original request. Furthermore, the agent does not see the entire state of the webpage; it only has access to the specific information and actions the website explicitly shares. This allows the website to protect potentially sensitive user data that has been entered on the page but is not meant to be shared with a third-party Inference Provider. As a final layer of control, the user can configure the Browser Agent to disable specific contexts they do not want the agent to see. This multi-level system of permissions ensures that interactions are both powerful and privacy-preserving.

This approach stands in contrast with two prevailing models of web-agent integration, each of which centralizes control at the expense of a key stakeholder. In one model, the website implements its own bespoke LLM support, forcing the user into a position where they must trust the site operator with their conversational data, which may be used in ways that are not transparent or aligned with the user's interests. In the opposing model (e.g., Claude for Chrome\footnote{https://www.anthropic.com/news/claude-for-chrome, accessed \today}, Perplexity Comet\footnote{https://www.perplexity.ai/comet/, accessed \today}), an inference provider deploys a universal agent that attempts to infer actions and state from a website's raw HTML and screenshots. This disempowers the website developer, who loses control over both the user experience and data privacy, as the agent may perform unintended actions or access data not meant for exposure. VOIX, by design, avoids this  by creating a standardized, explicit contract that balances the needs of both the user and the website developer.

\subsection{Reference Implementation}
We contribute a reference implementation of VOIX with a Chrome extension that enables chat and voice interaction with websites. Its operation begins with a script, injected into the website, which discovers all declared \texttt{<tool>} and \texttt{<context>} elements. This script also actively monitors the DOM for any changes, allowing it to maintain a dynamic and up-to-date model of the available tool and context space. \Autoref{fig:example-gui, fig:embedded-voix-elements} showcase an example task list application that embeds these VOIX elements alongside its normal HTML code.

The primary user interface is a side panel that supports multiple modes of interaction, see \autoref{fig:voix-sidepanel}. Users can communicate via standard text chat or through advanced voice inputs. This includes a voice mode that uses an on-device voice activity detection model to enable continuous conversation, as well as a transcription voice input that allows the user to see and edit their spoken words before sending them to the agent. To enhance discoverability, the side panel visualizes the available tools and contexts from the website and generates example prompts derived from them. It also allows the user to toggle model-specific features, such as \textit{thinking mode}.

Finally, in an options page of the chrome extension, the user can configure their preferred Inference Provider for both the language model and the transcription service, with support for any OpenAI-compatible API endpoint. This ensures that users retain full control over their data, their choice of models, and their costs.

The agentic LLM interaction in the extension maintains a conversation history. When the user makes a request, the contexts are prepended to the message, and a request is sent to the configured Inference Provider, including the currently available tools. When a tool call is generated, an event is triggered through the injected script to invoke the website's tool handler. If the tool returns something, this is reported back to the LLM Agent, which finally writes a message explaining what happened.

\subsection{Developing VOIX-supported Websites}
Integrating VOIX into a web application involves two primary steps: first, declaratively exposing the application's state and capabilities through HTML, and second, connecting these declarations to the application's existing logic. Using a task management application as a running example (\autoref{fig:example-gui}), this section explains how to implement VOIX-supported web apps.

An agent must first understand the current state of the application to act effectively. The \texttt{<context>} element provides this information as simple plain text. In a task management application, the list of current tasks is exposed so the agent knows what can be acted upon. This context can be dynamically populated when the application's state changes. Next, the application declares what actions an agent can perform. 

The \texttt{<tool>} element defines an action, its purpose, and its parameters. The call event handler is used to link the tool invocation to application logic. When the agent invokes this tool, the associated JavaScript function is executed. The function receives the parameters gathered by the agent in the event's detail property. Beyond simply invoking actions, VOIX also implements a way to report back a tool call's outcome. This is essential for confirming success, handling errors gracefully, and enabling tools that fetch data. VOIX facilitates this as an asynchronous event-driven path: A developer can signal that a tool will provide feedback by adding \texttt{return} attribute to its definition. When this attribute is present, the Agent waits and listens for a return event after the tool is called, before sending its next response to the user. As illustrated in \autoref{fig:embedded-voix-elements}, this event's detail payload contains a structured object indicating the result.

%% file: sections/hackathon.tex
\section{Hackathon}
\label{sec:hackathon}
To study the learnability and expressiveness of VOIX, we conducted an IRB-approved hackathon study spanning three days. In teams of two or three, 16 participants with varying degrees of web development experience implemented six VOIX-supported web apps for different use cases. This section explains the study setup, analyzes the resulting applications, and discusses the outcome to answer the following research questions:

\begin{enumerate}
    \item[Q1] Can developers learn to use VOIX tools within short time periods to create meaningful multimodal interactions in their web apps?
    \item[Q2] Is VOIX expressive enough to implement the interactions developers come up with in their use cases?
\end{enumerate}

\subsection{Study Setup}
To address our research questions, we conducted a three-day, in-person hackathon study. 
We recruited 16 participants (4 teams of three, 2 teams of two) who completed the study and submitted a final project. The event was held on-site, with complimentary catering and a required minimum of four hours of presence to incentivize participation and collaboration.

Participants came from diverse technical backgrounds, ensuring a representative sample of web developers. A pre-hackathon survey revealed a balanced distribution of self-reported web development experience: 18\% as advanced, 41\% identified as solid, 29\% as having basics, and 12\% as having none. Four of the six teams registered together, while the remaining two teams were formed during the kickoff session. 

The hackathon began with an onboarding session on Day 1, which included a study briefing, a hands-on VOIX workshop, and the announcement of the theme. The second day was dedicated entirely to project development, with facilitators providing support through observational mentorship. The event culminated on Day 3 with a final coding session, followed by live demos, and an awards ceremony in the afternoon.

The central theme for the hackathon was \textit{content creation}. This theme was intentionally chosen to allow teams the freedom to explore diverse and creative use cases for VOIX while requiring a complex workflow. To support their work, we provided each team with a comprehensive documentation package for VOIX, including code examples, and a set of API keys to give their VOIX Chrome extension access to a state-of-the-art LLM (\textit{Qwen3-235B-A22B} \cite{yang_qwen3_2025}) and transcription endpoint.

\begin{figure*}[tb]
    \centering
    \includegraphics[width=1\linewidth]{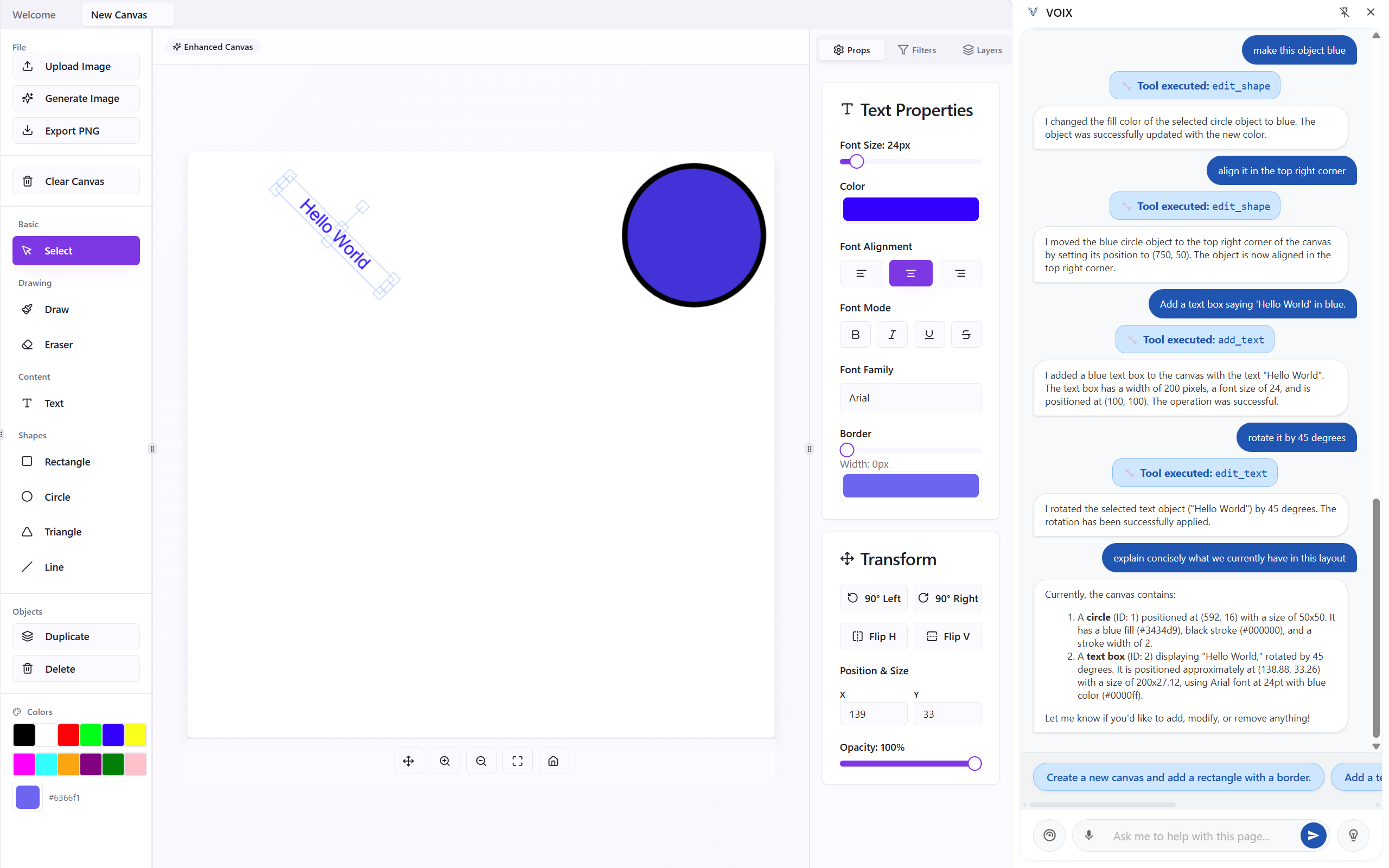}
    \caption{The graphic design application demonstrates synergistic multimodal interaction using VOIX: dynamic context elements contain information about the objects on the canvas and their state, allowing the Agent to understand \textit{which} objects to change \textit{how} in order to implement the users instructions. Then, a broad set of tools enables the LLM to create, edit, rearrange and delete objects.}
    \label{fig:creative-studio}
\end{figure*}

\subsection{Data Collection}
We employed a mixed-methods approach to gather comprehensive data on VOIX's learnability, expressiveness, and the challenges of its implementation.

\subsubsection*{Survey} Participants completed a post-hackathon survey to observe their understanding and perception of VOIX. Specifically, we collected the System Usability Scale \cite{brooke_sus_1996} and the Trust of Automated Systems Test (TOAST) \cite{wojton_initial_2020}.

\subsubsection*{Observational Mentorship} Throughout Facilitators engaged in casual conversations with participants and observed their workflows during the hackathon, capturing spontaneous reactions, collaboration patterns, and contextual challenges.

\subsubsection*{Interviews} Each team participated in a scheduled one-hour interview on one of the three days, allowing for in-depth discussion of their design process, implementation hurdles, and experience with the framework.

\subsubsection*{Project Artifacts} We collected the final source code of all six projects for a detailed analysis of the implemented VOIX integrations.\\

\noindent To motivate participants, we offered a tiered prize pool. Members of the winning team each received EUR 300, with prizes of EUR 200, EUR 100, and EUR 50 per member for the second, third, and remaining teams, respectively.

\subsection{Expressiveness of the Applications}
Development teams built a diverse set of VOIX-supported web applications. The projects included a Creative Studio for graphic design, a Fitness App for generating workout plans, a Soundscape Creator for designing ambient audio environments, a Kanban-style Project Management tool, an Anki Creator for building flashcard decks, and a form-based fantasy Character Creation website. Detailed descriptions and screenshots of these applications can be found in \autoref{sec:appendix-apps}. An analysis of these applications demonstrates that our framework successfully meets two of its primary design goals: it is sufficiently expressive to support sophisticated multimodal interactions (\ref{req:expressive}), and it enables practical, effective scoping using standard web development patterns (\ref{req:representation}).

The applications confirmed the framework's expressiveness by demonstrating its capacity to handle a range of interaction patterns. The framework's ability to process abstract, high-level user inputs was evident in the Fitness App, where a command such as "create a full high-intensity training plan for my back and shoulders" was successfully interpreted and executed, selecting the appropriate exercises, set counts, and repetition schemes from its knowledge base to assemble and display the complete routine. Similarly, the Soundscape Creator allowed users to make broad, conceptual requests like "make it sound like a rainforest." Furthermore, participants implemented both complementary and redundant multimodal actions. The Creative Studio offered a clear example of complementary input, supporting deictic interactions where a user could click on a canvas element while issuing a voice command like "rotate this by 45 degrees" (see \autoref{fig:creative-studio}). This showcased the framework's ability to fuse inputs from different modalities to resolve a single intention. In contrast, applications like the Project Management tool provided redundant modalities, allowing users to perform the same action, such as creating a task, through either the graphical interface or an equivalent voice command.

\hyperref[req:representation]{Requirement \ref*{req:representation}} for the framework was to allow developers to scope the available tools and context to prevent ambiguity and enable dynamic, stateful interactions. The hackathon participants demonstrated that this can be achieved idiomatically using the component-based architectures of modern JavaScript frameworks. Developers naturally scoped multimodal functionality by defining the available tools and context within their React or Vue components, thus restricting these capabilities to the specific UI region where a component was active. This pattern of context-sensitive availability was common. For instance, the Anki Creator only exposed the tool for image generation when the user was interacting with a specific flashcard category component, while the Project Management tool only made the create task tool available when the main Kanban board was rendered. This scoping was often dynamic, changing in response to the application's state. The Creative Studio and Fitness App, for example, used conditional rendering logic to expose editing tools only after a user had selected a specific element. These implementations confirm that developers can use familiar, state-driven patterns from component-based frameworks to effectively manage the scope of multimodal interactions, making the integration of VOIX both powerful and practical.

\subsection{Learnability of VOIX}
To evaluate how easily developers could learn and apply the VOIX framework (Q1), we gathered quantitative data using standardized usability questionnaires and analyzed the practical outcomes of the hackathon. Our primary measure was the System Usability Scale (SUS) \cite{brooke_sus_1996}, a widely adopted and reliable tool for assessing perceived usability. The results from the post-hackathon survey (N=16) yielded a mean SUS score of 72.34 (SD=14.82). A score above the industry average of 68 is considered "\textit{good}". This score indicates that, on the whole, developers found the VOIX framework to be usable and easy to learn.

In addition to usability, we assessed participants’ trust in VOIX using the Trust of Automated Systems Test (TOAST) \cite{wojton_initial_2020}. Following the validated two-factor structure of the scale, we report results separately for \textit{System Understanding} and \textit{System Performance}. The mean score for System Understanding was 5.81 (SD = 0.85), while the mean score for System Performance was 5.14 (SD = 0.87), both on a 1--7 scale. These results suggest that participants not only developed a solid conceptual grasp of VOIX’s functionality, but also perceived it as performing reliably. Together with the SUS results, these findings provide evidence that VOIX is both learnable and capable of inspiring confidence in its expressiveness and reliability.

The interviews and observations reinforced that VOIX was easy to learn and apply within the limited timeframe. Participants emphasized that the framework’s reliance on only two additional HTML tags (\texttt{<content>} and \texttt{<tool>}) allowed them to integrate functionality as naturally as defining a \texttt{<button>} element. Some teams reported that this simplicity fit well with familiar patterns in frameworks like React and Vue, where tools and context could be scoped through standard component logic and conditional rendering. Several teams noted that this alignment simplifies the process and made rapid prototyping feasible.

At the same time, participants highlighted a conceptual challenge: deciding which tools were meaningful to implement. While technically straightforward, teams debated whether specific actions provided real benefit compared to existing UI interactions, sometimes defaulting to simple CRUD-style tools. Thus, while the mechanics of VOIX were quickly mastered, identifying valuable use cases required more reflection.

%% file: sections/discussion.tex
\section{Discussion}
The architectural design of VOIX and the empirical findings from our hackathon study confirm that the framework successfully meets its core design requirements. The results demonstrate that by embedding machine-readable affordances directly into the DOM, VOIX offers a practical path toward a more robust, decentralized, and privacy-preserving Agentic Web.

The framework's architecture directly satisfies the requirement for Privacy and Safety (\ref{req:privacy}). By design, VOIX creates a clear trust boundary that separates the user's conversational data from the website's domain. The Browser Agent sends user prompts directly to the user's chosen LLM provider, ensuring the website operator never sees the content of the conversation. Conversely, the website only exposes explicitly declared information via the context tag, protecting sensitive or proprietary user data from being indiscriminately scraped by a third-party agent.

The study also verified the need for an Optimal Representation (\ref{req:representation}) of a site's capabilities. The tool and context primitives provide a concise, structured summary of affordances, eliminating the inefficiency and brittleness of parsing entire DOMs. The hackathon results provided strong empirical validation for this principle, as development teams naturally leveraged component-based frameworks like React and Vue to dynamically scope these affordances. This common pattern ensured that tools were only discoverable when relevant, inherently linking the agent's capabilities to the real-time state of the user interface.

Furthermore, VOIX is architecturally designed to be Efficient to Host (\ref{req:hosting}). The framework is fundamentally client-side, placing the computational and financial burden of LLM inference on the user's agent and chosen provider, not the website owner. This decentralization removes a significant barrier to entry and aligns with the web's ethos, making the widespread adoption of the standard feasible for developers and organizations of all sizes.

The framework's success in the hackathon underscored its adherence to being Standardized and Developer-Friendly (\ref{req:standardized}). VOIX is built upon familiar web patterns: simple, declarative HTML tags and standard JavaScript event listeners. The study provided direct evidence of its learnability, as all teams were able to create functional, agent-enabled web applications within the short three-day timeframe. Also, the participants reported that the framework was easy to grasp, with participants emphasizing that its two additional tags could be used as naturally as standard HTML elements.

Finally, the applications developed during the hackathon confirmed that the framework is sufficiently Expressive (\ref{req:expressive}) to support the key patterns of effective multimodal interaction. The projects demonstrated a range of sophisticated patterns, including abstract, high-level commands, complementary deictic interactions where a click was paired with a voice command, and redundant modalities that allowed users to perform the same action through either the GUI or voice.

\begin{table}[h]
\centering
\caption{Latency Benchmark comparing VOIX with Perplexity Comet\tablefootnote{\url{https://www.perplexity.ai/comet}, accessed \today} and BrowserGym \cite{chezelles_browsergym_2025} with GPT-5-mini. All latency values are in seconds. Details in \autoref{sec:appendix-latency}.}
\label{tab:latency-benchmark}

\begin{tabular}{@{}lccc@{}}
\toprule
\textbf{Task} & \textbf{VOIX} & \textbf{Comet} & \textbf{BrowserGym} \\ 
\midrule

\multicolumn{4}{@{}l}{\textbf{Creative Studio}} \\ 
Add a blue triangle & 2.32 & 27.21 & 25.29 \\
Rotate the green triangle 90° & 1.11 & 89.12 & \textit{Failed} \\
Delete selected object & 0.96 & 16.29 & 5.69 \\
Export as an image & 1.30 & 10.12 & 4.25 \\
\midrule

\multicolumn{4}{@{}l}{\textbf{Fitness App}} \\
Create a full week HIIT plan & 14.38 & 229.52 & 1271.00 \\
Start Day 1 of plan & 1.07 & \textit{Failed} & 26.27 \\
Export Day 2 \& 5 as PDF & 1.87 & 17.37 & 13.82 \\
Show workout statistics & 0.91 & 10.42 & 5.79 \\
\midrule

\multicolumn{4}{@{}l}{\textbf{Project Management Tool}} \\
Create task & 1.62 & 26.14 & 33.01 \\
Report on tasks in progress & 1.30 & 4.43 & 6.90 \\
Copy task to another project & 2.67 & 61.94 & \textit{Failed} \\
\bottomrule
\end{tabular}
\end{table}
\subsection{Latency Comparison}
A latency comparison (\autoref{tab:latency-benchmark}) demonstrates a fundamental architectural failure in the industry standard affordance inference based paradigm. As established by foundational research, the classic threshold for an action to be perceived as "instantaneous" is 100-200ms \cite{forch_are_2017}. However, for the direct manipulation tasks often combined with speech in multimodal commands, performance is measurably degraded when latency exceeds as little as 25ms \cite{deber_how_2015}. Our tests show that vision-based agents exhibit latencies ranging from 4.25 seconds to over 21 minutes for complex tasks with long-horizon multi-step action chains. These response times are not merely slow; they are orders of magnitude beyond these critical perceptual and performance thresholds. This inherent architectural latency makes it impossible to create the tight temporal binding between modalities, which is the primary cue used to fuse speech and gesture into a single, coherent perceptual event \cite{oviatt_perceptual_2000}. In comparison, VOIX enables efficient, single-step solutions to these tasks since feedback is immediate and long loops of inferring affordances, executing actions and checking success can be avoided.

\subsection{Limitations and Future Work}
While the hackathon study demonstrated VOIX's promise, its controlled environment does not fully capture the challenges of real-world adoption. A primary limitation is the long-term developer burden, which manifests in two key areas: maintenance and conceptual design. First, integrating VOIX into large, legacy codebases introduces a risk where the declarative tools fall out of sync with the evolving graphical user interface. If a new GUI feature is added without a corresponding VOIX implementation, the multimodal experience becomes inconsistent and frustrating, undermining the framework's goal of seamless interaction. Preventing this requires significant development discipline and new testing methodologies not explored in our short-term study.

Second, VOIX demands a significant conceptual shift for developers, moving from a familiar visual paradigm to an affordance-centric one. This creates a difficult abstraction dilemma when designing tools. On one hand, low-level tools that mirror atomic GUI actions are simple to maintain but offer little performance benefit over traditional agents. On the other hand, high-level, intent-aligned tools provide high performance but are difficult to design comprehensively and become brittle when a user's goal falls outside the developer's preconceived notions. Finding a balanced and effective middle ground imposes a substantial design burden on development teams, representing a key hurdle for practical adoption.

%% file: sections/conclusion.tex
\section{Conclusion}
This work introduced VOIX, a web-native framework designed to resolve the fundamental misalignment between AI agents and the human-first web. By enabling websites to expose reliable, auditable, and privacy-preserving affordances through simple declarative HTML, VOIX offers a concrete solution to the brittleness, inefficiency, and security risks inherent in current screen-scraping and DOM-parsing approaches.

Our empirical evaluation through a multi-team hackathon demonstrated that VOIX is a practical and accessible tool for developers. The study confirmed that developers, regardless of prior experience, could rapidly build functional, agent-enabled web applications. Furthermore, the analysis of the resulting applications revealed the emergence of sophisticated, synergistic multimodal interaction patterns, validating that VOIX is expressive enough to realize the benefits of combining direct manipulation and natural language as envisioned by foundational human-computer interaction research.

By shifting the responsibility of defining agent capabilities from the agent provider to the website developer, VOIX creates a balanced, decentralized architecture that respects user privacy and developer control. It represents a critical and deployable step toward an Agentic Web built for seamless and secure collaboration between humans and AI.

%% file: appendix/index.tex
\appendix
\input{appendix/A}
\newpage
\input{appendix/B}

%% file: appendix/A.tex
\newpage
\NewDocumentEnvironment{app}{O{} O{}}{%
  \smallskip\noindent\includegraphics[width=\linewidth]{#2}\vspace{4pt}
  \par\noindent\textbf{#1.}
}{

}
\section{Hackathon Applications}
\label{sec:appendix-apps}

\begin{app}[Soundscape Creator][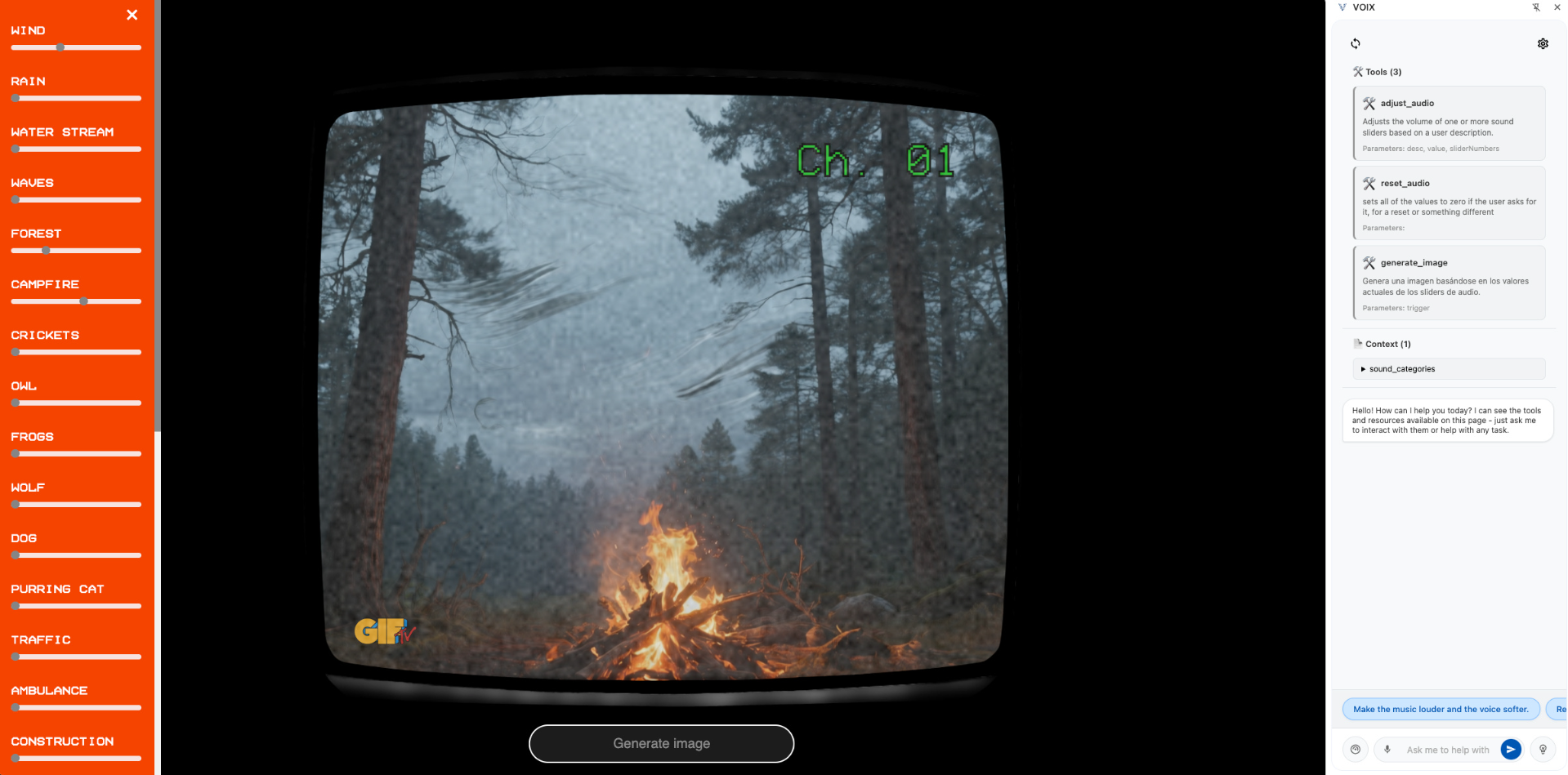]
This vanilla HTML application allows users to create soundscapes by mixing samples on a set of sliders. Meanwhile, the application generates images that match the current soundscape using a diffusion model. The application implements VOIX tools to change the soundscape based on user instructions, potentially changing multiples sliders at once.\\[4pt]
\end{app}

\begin{app}[Fitness App][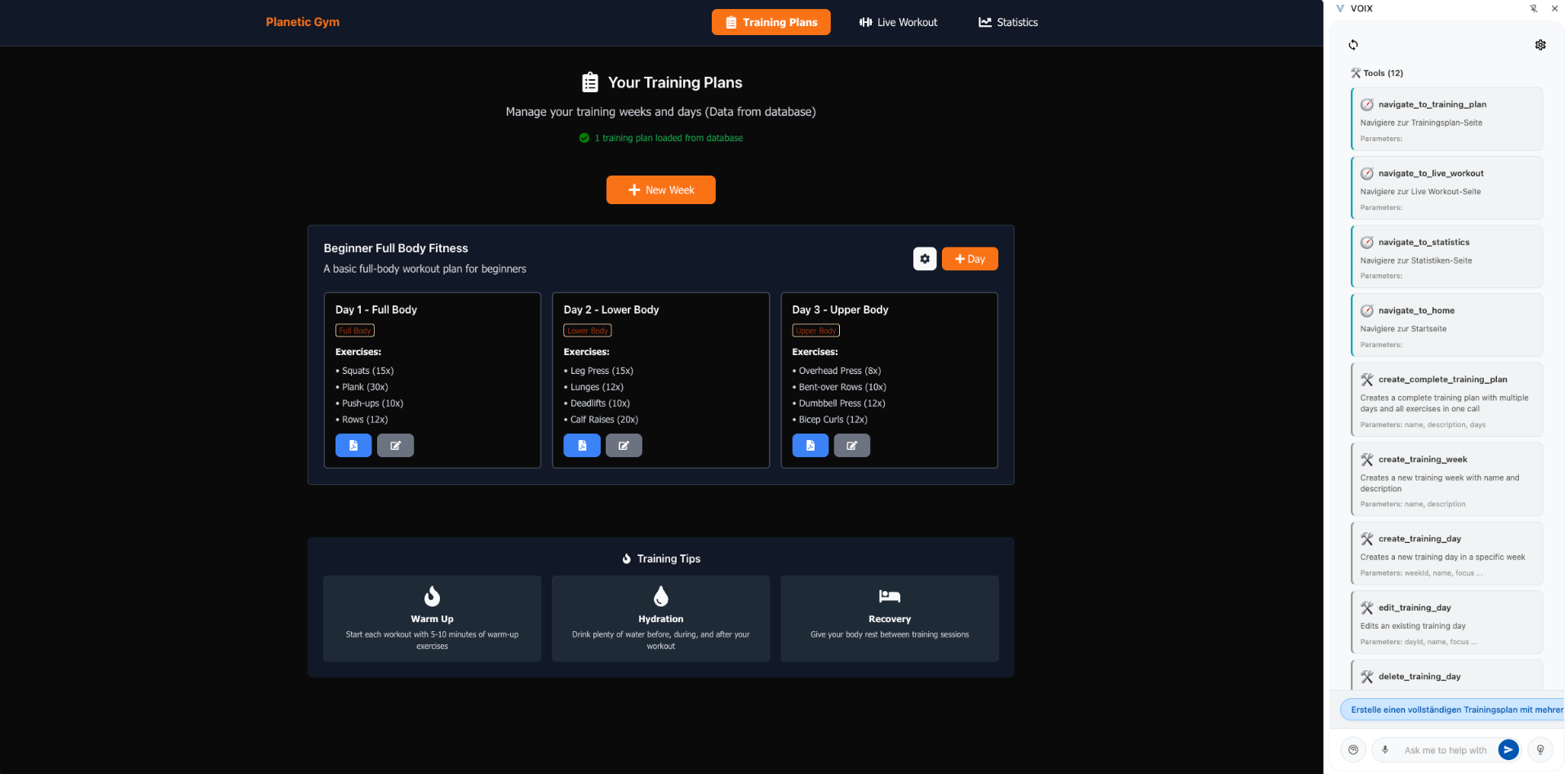]
This React app is a fitness application for creating multi-week training plans with customizable days and exercises. It includes an interactive workout interface with timers, set tracking, and real-time exercise management. Performance is tracked through data dashboards with an option to export reports as PDFs. The app integrates VOIX tools to operate core functions hands-free and enable LLM-based exercise planning and support.\\[4pt]
\end{app}

\begin{app}[Character Creator][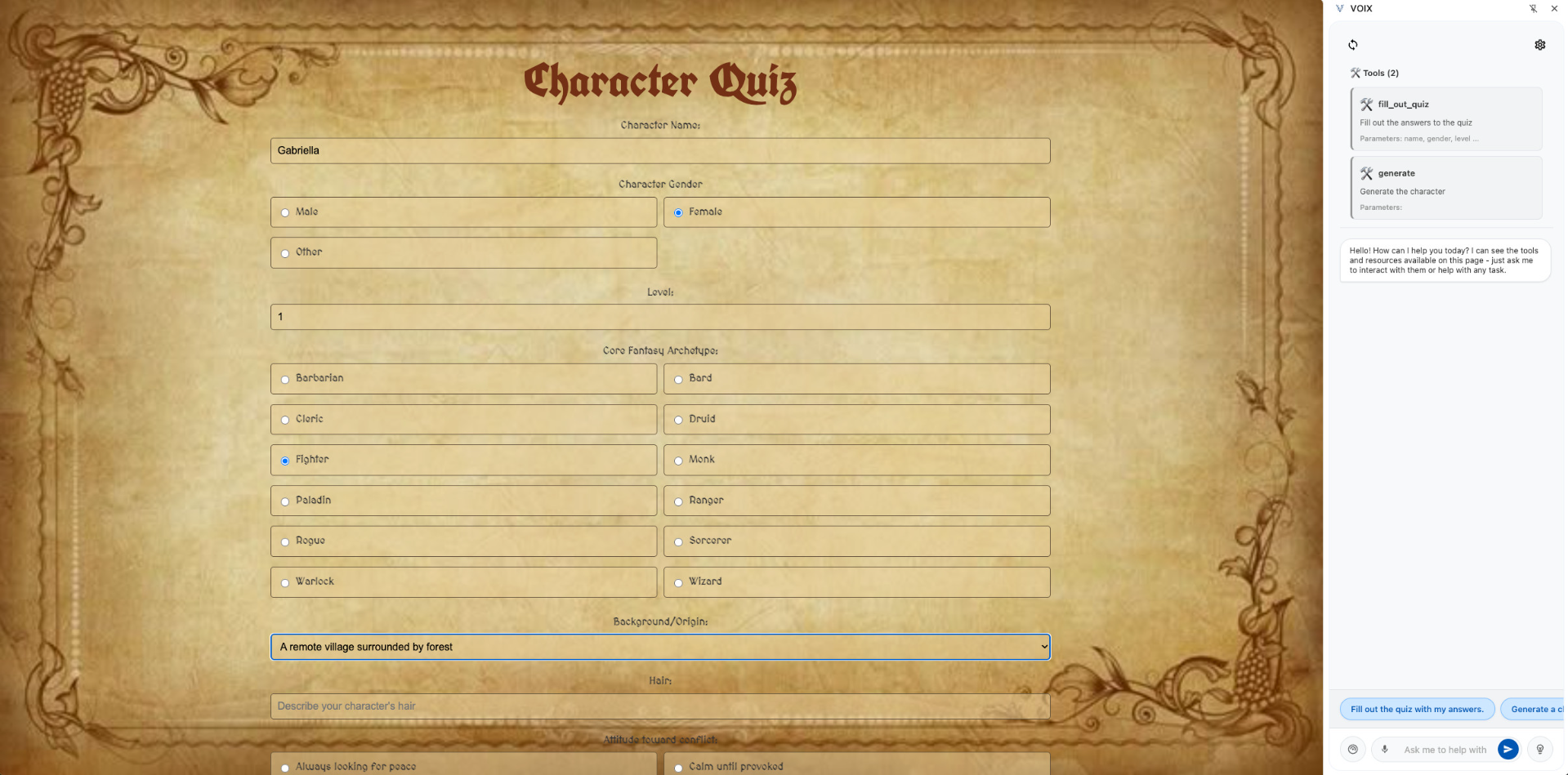]
This vanilla HTML website helps users create fantasy characters using a quiz and facilitates LLM-guided role playing with those characters. It primarily uses VOIX tools to support filling out the quiz using instructions instead of tediously selecting each question one-by-one.\\[4pt]
\end{app}

\begin{app}[Creative Studio][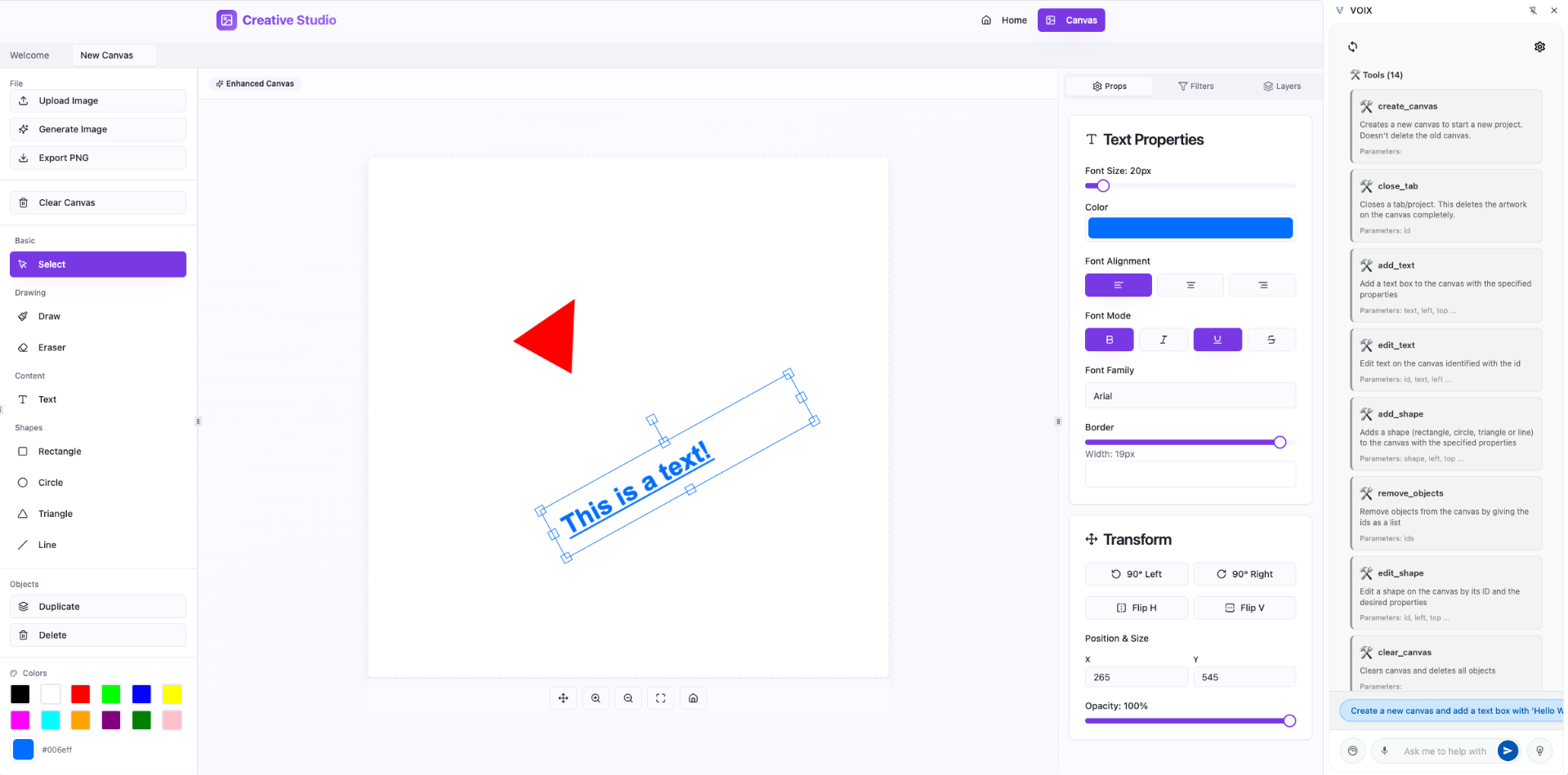]
With this React app, users can create visual designs on a canvas using basic shapes and text. The interface provides tools to modify object properties like color, size, rotation, and position. The application exposes its core functionalities as VOIX tools, allowing users to perform multimodal interactions combining selection and natural language instructions to perform tasks.\\[4pt]
\end{app}

\begin{app}[Project Management Tool][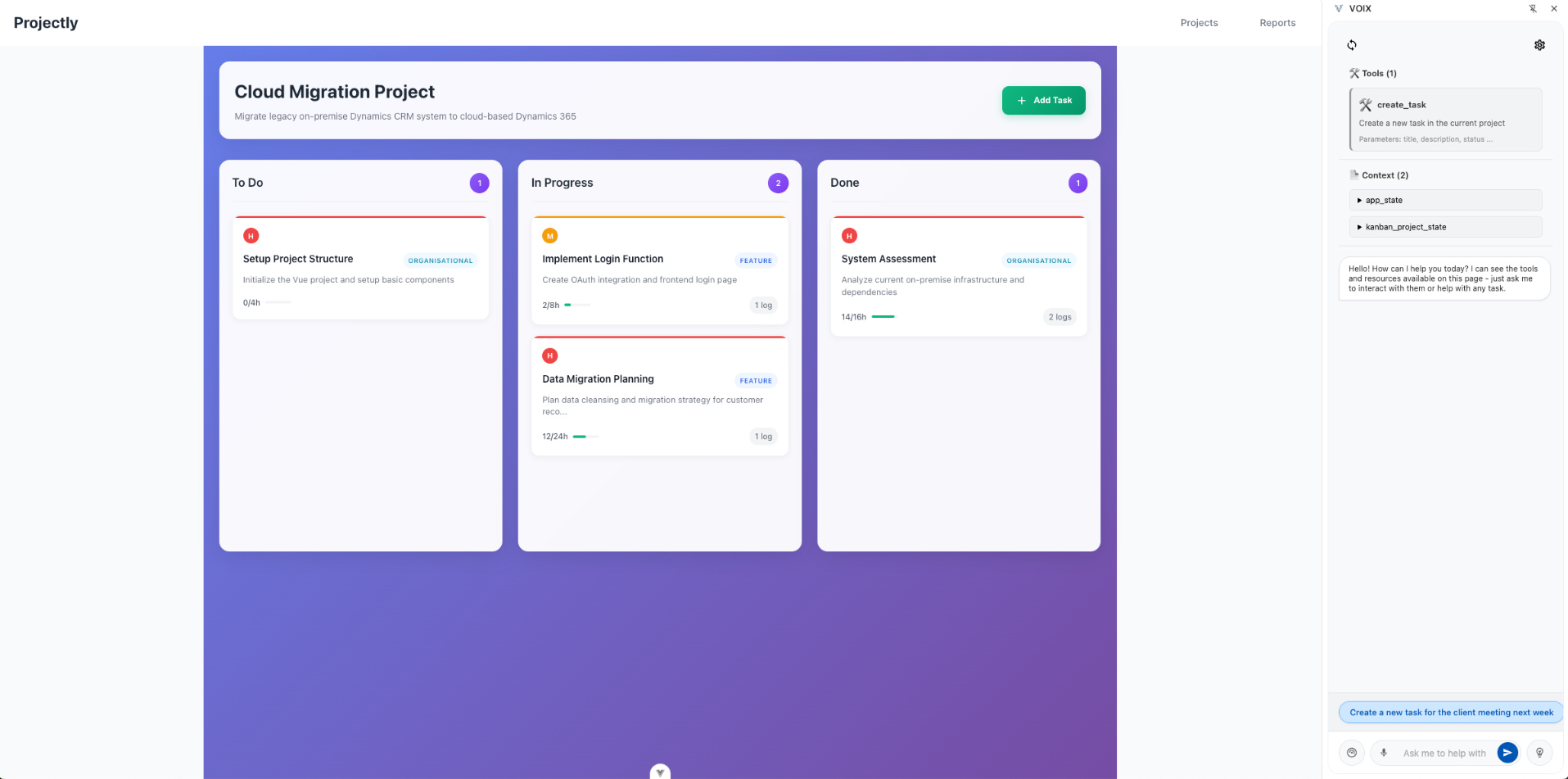]
This is a Vue.js-based project management tool that uses a Kanban board to track tasks through different stages, such as 'To Do', 'In Progress', and 'Done'. Through the VOIX interface, users can interact with the board using natural language to create new tasks.\\[4pt]
\end{app}

\begin{app}[Anki Creator][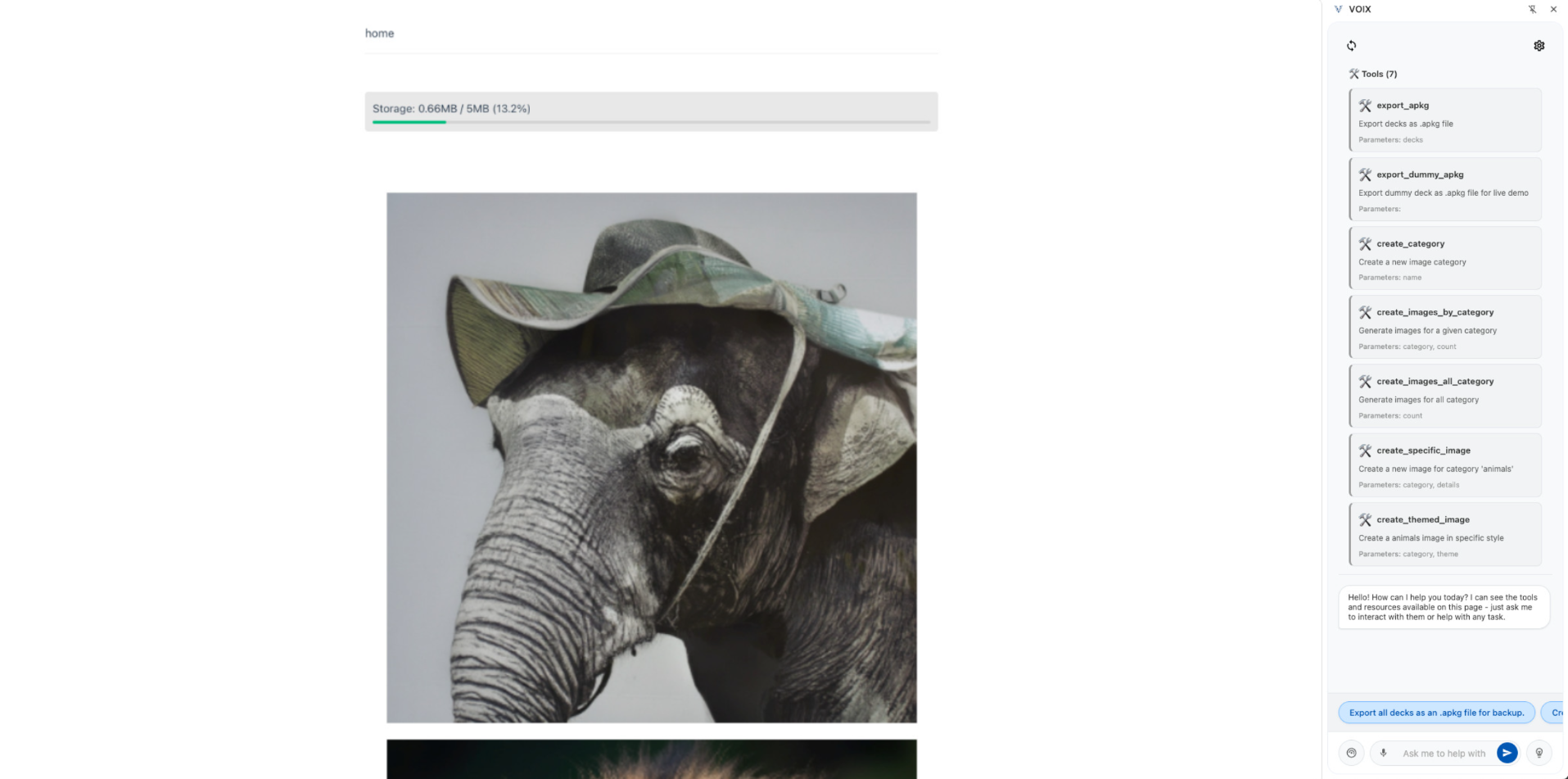]
This Vue.js-based application is a tool for generating image-based flashcard decks compatible with the Anki spaced repetition system. Users can define categories and then use AI-powered tools to generate images for those categories. The entire workflow from creating categories and generating images to exporting the final deck is controlled via natural language commands in the VOIX interface.
\end{app}

%% file: appendix/B.tex
\section{Latency Benchmark}
\label{sec:appendix-latency}
This appendix details the methodology used for the latency benchmark comparison presented in \autoref{tab:latency-benchmark} and Section 5.1 of the main paper.

\subsection{Benchmark Objective}
The primary objective of this benchmark was to quantitatively compare the end-to-end task completion latency of our proposed \textbf{VOIX framework} against two representative \textbf{affordance inference-based agentic systems}: Perplexity Comet and BrowserGym. The goal was to measure the performance difference between a declarative, machine-readable approach (VOIX) and approaches that rely on interpreting visual UIs and raw DOM structures.

\subsection{Systems Under Test}
Three distinct systems were evaluated:
\begin{itemize}[leftmargin=*]
    \item \textbf{VOIX}: The web applications built by hackathon participants using the VOIX framework, controlled via our reference Chrome extension connected to the Qwen3-235B-A22B large language model.
    \item \textbf{Perplexity Comet}: A commercial, vision-based web agent that infers actions from screenshots and HTML. Tests were conducted using the publicly available version as of September 4, 2025.
    \item \textbf{BrowserGym}: An open-source research framework for web agents \cite{chezelles_browsergym_2025}. We used it with a GPT-5-mini model to execute tasks on the same web applications used for the VOIX tests. The BrowserGym agent was provided with the same high-level objective for each task.
\end{itemize}

\subsection{Measurement Protocol}
\begin{itemize}[leftmargin=*]
    \item \textbf{Latency Measurement}: For each task, latency was measured in seconds from the moment the natural language prompt was submitted by the user to the moment the requested change was fully rendered and visually confirmed on the screen.
    \item \textbf{Trials}: Each task was executed \textbf{once per platform} to measure a direct, single-shot completion time. If a task did not complete successfully, it was retried up to a maximum of three total attempts. The VOIX framework never required a retry for any task.
    \item \textbf{Failures}: A task was marked as \textit{Failed} if the agent could not complete it within the three-attempt limit or exceeded a timeout of 25 minutes.
\end{itemize}

\subsection{Task Descriptions}
The following lists the exact natural language prompts used for each task.

\subsubsection*{\textbf{Creative Studio}}
\begin{itemize}[leftmargin=*]
    \item \textbf{Add a blue triangle}: The agent was instructed with the prompt: \textit{"add a blue triangle to the canvas"}.
    \item \textbf{Rotate the green triangle 90°}: The agent was instructed with the prompt: \textit{"rotate the green triangle by 90 degrees to the left"}.
    \item \textbf{Delete selected object}: With an object already selected, the agent was instructed with the prompt: \textit{"delete this object"}.
    \item \textbf{Export as an image}: The agent was instructed with the prompt: \textit{"export this as as an image"}.
\end{itemize}

\subsubsection*{\textbf{Fitness App}}
\begin{itemize}[leftmargin=*]
    \item \textbf{Create a full week HIIT plan}: The agent was given the high-level prompt: \textit{"create a full week high-intensity training plan for my back and shoulders"}.
    \item \textbf{Start Day 1 of plan}: The agent was instructed with the prompt: \textit{"start day one of my high intensity training plan"}.
    \item \textbf{Export Day 2 \& 5 as PDF}: The agent was instructed with the prompt: \textit{"export day 2 and 5 from my training plan as pdf"}.
    \item \textbf{Show workout statistics}: The agent was instructed with the prompt: \textit{"show me statistics on my workout routine"}.
\end{itemize}

\subsubsection*{\textbf{Project Management Tool}}
\begin{itemize}[leftmargin=*]
    \item \textbf{Create task}: The agent was instructed with the prompt: \textit{"Create a task to finish the database optimization by wednesday"}.
    \item \textbf{Report on tasks in progress}: The agent was instructed with the prompt: \textit{"Give me a report of how many tasks are currently in progress on the ecommerce platform project"}.
    \item \textbf{Copy task to another project}: The agent was instructed with the prompt: \textit{"copy the most recently added task from the Website Redesign project over to this one"}.
\end{itemize}